\documentclass{article}
\usepackage{ijcai}
\usepackage{times}
\usepackage{soul}
\usepackage[hidelinks]{hyperref}
\usepackage[utf8]{inputenc}
\usepackage[small]{caption}
\usepackage{graphicx}
\usepackage{amsmath}
\allowdisplaybreaks[4]
\usepackage{amsthm}
\usepackage{booktabs}
\usepackage{algorithm}
\usepackage{algorithmic}
\usepackage[switch]{lineno}
\usepackage[T1]{fontenc}
\usepackage{times}
\usepackage{soul}
\usepackage{multirow} 
\usepackage{amsfonts}
\usepackage{amssymb}
\usepackage{mathrsfs}
\usepackage{boldline,multirow}
\usepackage{makecell}
\usepackage{epsfig}
\usepackage{dsfont}
\usepackage{color}
\usepackage{siunitx}
\usepackage{colortbl}
\usepackage{cases}
\usepackage{bbding}
\usepackage{url}
\definecolor{gbypink}{rgb}{0.99, 0.91, 0.95}

\newcommand{\ve}[1]{\mathbf{#1}}
\newtheorem{theorem}{Theorem}[section]

\newtheorem{lemma}[theorem]{Lemma}

\begin{document}
%
%\title{DPMs-SDE: A Novel 2D Segmentation Dataset Expansion Technique Using Diffusion Models}
\title{DiffuseExpand: Expanding Dataset for 2D Medical Image Segmentation Using Diffusion Models}

% \author{
%     Author Name
%     \affiliations
%     Affiliation
%     \emails
%     email@example.com
% }

\author{
Shitong Shao$^1$
\and
Xiaohan Yuan$^1$\and
Zhen Huang$^2$\and
Ziming Qiu$^1$\and
Shuai Wang$^3$\and
Kevin Zhou$^2$\footnote{Corresponding author}
\affiliations
$^1$Southeast University, Nanjing, China\\
$^2$Univeristy of Science and Technology of China, Hefei, China\\
$^3$Tsinghua University, Beijing, China \\
\emails
\{1090784053sst,s.kevin.zhou\}@gmail.com, xiaohan\_yuan@163.com, zhenhuang@mail.ustc.edu.cn, qiuziming@seu.edu.cn, s-wang20@mails.tsinghua.edu.cn}
% \email{\{1090784053sst,s.kevin.zhou\}@gmail.com, xiaohan_yuan@163.com, zhenhuang@mail.ustc.edu.cn, qiuziming@seu.edu.cn, s-wang20@mails.tsinghua.edu.cn}
\maketitle    
\begin{abstract}
Dataset expansion can effectively alleviate the problem of data scarcity for medical image segmentation, due to privacy concerns and labeling difficulties. However, existing expansion algorithms still face great challenges due to their inability of guaranteeing the diversity of synthesized images with paired segmentation masks. In recent years, Diffusion Probabilistic Models (DPMs) have shown powerful image synthesis performance, even better than Generative Adversarial Networks. Based on this insight, we propose an approach called DiffuseExpand for expanding datasets for 2D medical image segmentation using DPM, which first samples a variety of masks from Gaussian noise to ensure the diversity, and then synthesizes images to ensure the alignment of images and masks. After that, DiffuseExpand chooses high-quality samples to further enhance the effectiveness of data expansion. Our comparison and ablation experiments on COVID-19 and CGMH Pelvis datasets demonstrate the effectiveness of DiffuseExpand. {\textit{Our code is released at \url{https://github.com/shaoshitong/DiffuseExpand}.}}
\end{abstract}
\section{Introduction}
Image segmentation facilitates clinical indexes for diagnosis and treatment by delineating areas of interest, providing the basis for medical image analysis~\cite{wang2022medical}. Due to high privacy of medical images and difficulty of data annotation, existing labeled datasets typically have limited scale, thereby leading to the challenge of training segmentation models with high accuracy and strong generalization ability~\cite{privacy_medical,public_privacy}. To overcome this, data expansion~\cite{compare_1,compre_2}, which consists of both manual data augmentation and sample synthesis, is studied to effectively improve the accuracy of data-driven methods. When the generative model is excellent enough, sample synthesis can obtain more diverse and hard samples compared with manual data augmentation~\cite{anaya2022fourier}. 

\begin{figure*}[t]
\includegraphics[width=0.95\textwidth,trim={0cm 0cm 0cm 0cm},clip]{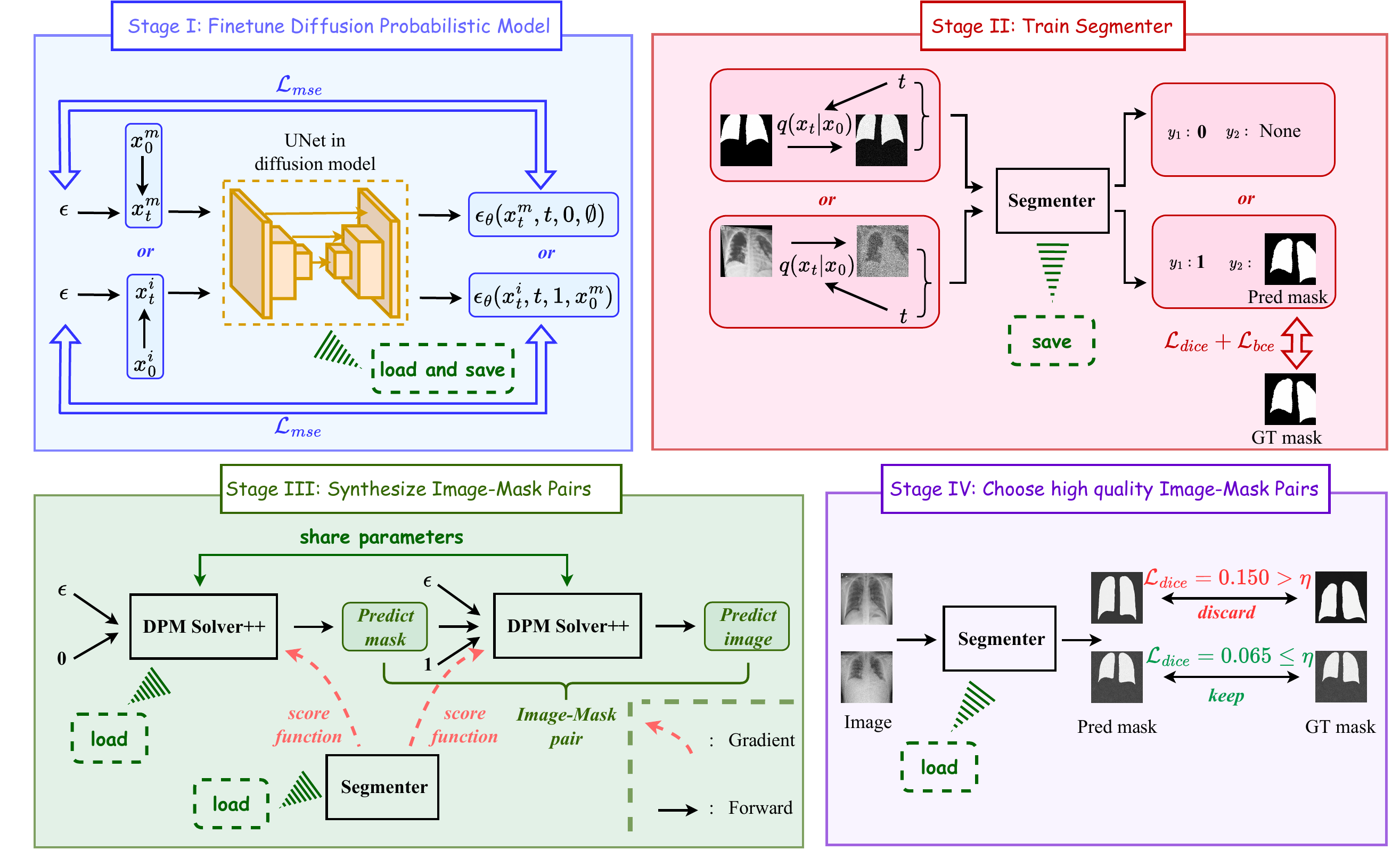}
\caption{The overall framework of DiffuseExpand, which contains four progressive stages.}
\label{fig:overall_structure}
\vspace{-5pt}
\end{figure*}
Generative models are primarily represented by Generative Adversarial Networks (GANs) and their variants~\cite{Goodfellow2014GenerativeAN,mirza2014conditional,zhu2017unpaired}. Although these models have shown remarkable performance in various tasks, they are not without limitations. Specifically, they often suffer from problems such as mode collapse and limited sampling diversity, which can limit their effectiveness~\cite{tronchin2021evaluating}. To overcome these limitations, researchers have proposed Diffusion Probabilistic Models (DPMs)~\cite{ddpm_begin}, inspired by thermodynamic Brownian motion. DPMs offer a more stable training process and have shown remarkable performance in synthesizing high-quality images. Furthermore, with the theories for accelerated sampling and conditional control in DPMs advancing, it opens up new possibilities for their application in semantic data expansion~\cite{adm2021,adm2022_medical,ddim,dpm_solver,dpm_solver++}.

Different from the single image synthesis task~\cite{Bermdez2018LearningIB,Madani2018SemisupervisedLW,Nie2018MedicalIS}, it is crucial and challenging to synthesize paired images and masks at the same time to build the expanded segmentation dataset. Inspired by~\cite{Wang_2022_ACCV} that synthesizes images via masks, we design DiffuseExpand using Diffusion Probabilistic Models, which first synthesizes masks from Gaussian noise and then utilizes the masks as conditions to synthesize corresponding images, thus enabling the pairing of image and mask and guaranteeing the sufficient variety in the synthetic sample pairs. In addition, we employ a well-trained model to discard bad samples to further ensure the quality of the expanded data. In summary, the contributions of this study can be
summarized as follows:
\begin{itemize}
    \item[$\bullet$] We introduce DPMs into segmentation data expansion and use their advantages to synthesize high-quality sample pairs. To the best of our knowledge, DPMs have yet to be deeply explored in the field of sample pair expansion. 
    \item[$\bullet$] We design an algorithm called DiffuseExpand to achieve efficient data expansion based on conditional bootstrapping of DPMs and the application of neural networks to discard bad samples.
    \item[$\bullet$] We evaluate our method on COVID-19 and CGMH Pelvis datasets~\cite{covid,cgmh}. The experiments show that DiffuseExpand yields expanded image-mask pairs with high fidelity and diversity, better than other methods.
\end{itemize}
\section{DPMs Background}
\label{sec:diffusion_model}
DPMs have become a powerful generative model in the field of computer vision~\cite{ddpm_begin,adm2021,adm2022_medical,poole2023dreamfusion}, which consists of a \textit{forward diffusion process} $q(x_t|x_0)$ and a \textit{reverse denoising process} $p_\theta(x_{t-1}|x_{t})$ at time $t\in \{1,2,\cdots,T\}$ with a \textit{noise prediction model} $\mu_\theta$. In training, DPMs gradually add Gaussian noise to sample $x_0$ based on the \textit{forward diffusion process}, such that $x_t = \alpha_t x_0 + \sigma_t \epsilon,\quad \epsilon\!\sim\!\mathcal{N}(\mathbf{0},\mathbf{I})$, where $\alpha_t$ is monotonically decreasing and $\sigma_t$ is monotonically increasing. Then DPMs align $\epsilon$ with the output $\mu_{\theta}(x_t,t)$ of the \textit{noise prediction model} as $\mathop{argmin}_{\theta} \mathbb{E} _{x_0,\epsilon,t}||\mu_\theta(x_t,t)-\epsilon||_2^2.$ And in sampling, the \textit{reverse denoising process} can be modeled as a reverse-time Stochastic Differential Equation (SDE) if we assume that the time variable $\hat{t}\in [0,1]$ is continuous~\cite{sde}. Let us define $f(\hat{t}) = \frac{d \log(\alpha_{\hat{t}})}{d\hat{t}}$ and $g(\hat{t}) =\sqrt{\frac{d\sigma_{\hat{t}}^2}{d\hat{t}}-2\frac{d \log \alpha_{\hat{t}}}{d{\hat{t}}}\sigma_{\hat{t}}^2}$~\cite{dpm_solver}, then we convert the discrete $q(x_t|x_{t-1})$ and $p_\theta(x_{t-1}|x_t)$ to the continuous \textit{forward diffusion process} and \textit{reverse denoising process} as follows
\begin{equation}
\small
\begin{aligned}
& q(x_t|x_{t-1}) =\!\!> dx_{\hat{t}} = f({\hat{t}})x_{\hat{t}}d{\hat{t}}+g({\hat{t}})dw_{\hat{t}},\\
& p_\theta(x_{t-1}|x_{t})=\!\!> dx_{\hat{t}} = [f({\hat{t}})-g^2({\hat{t}})\nabla_x\log q_{\hat{t}}(x_{\hat{t}})]x_{\hat{t}}d{\hat{t}}+g({\hat{t}})d\hat{w}_{\hat{t}}.\\
\end{aligned}
\label{eq:forward_and_reverse_process}
\end{equation}
where $w_{\hat{t}}$ and $\hat{w}_{\hat{t}}$ are the standard Wiener process, and $\nabla_x\log q_{\hat{t}}(x_{\hat{t}})$ is the \textit{score function}. Typically, past works~\cite{dpm_solver,dpm_solver++,adm2021} have used $\frac{\mu_\theta(x_{\hat{t}},{\hat{t}})}{-\sigma_{\hat{t}}}$ to estimate $\nabla_x\log q_{\hat{t}}(x_{\hat{t}})$. Subsequently, Song \emph{et al.} propose that SDE could be converted to a \textit{probability flow} Ordinary Differential Equation (ODE) via the Fokker-Planck equation~\cite{fpequation}, thus the \textit{reverse denoising process} becomes a deterministic process, \emph{i.e.}, $x_{b}$ is fixed when DPMs obtain $x_{b}$ by $x_a$~\cite{sde} ($0\leq b\leq a\leq 1$). The new \textit{probability flow} ODE can be formulated as
\begin{equation}
\begin{aligned}
& dx_{\hat{t}} = \left[f(\hat{t})x_{\hat{t}}+g^2({\hat{t}})\frac{\mu_\theta(x_{\hat{t}},\hat{t})}{2\sigma_{\hat{t}}}\right]d{\hat{t}}.\\
\end{aligned}
\label{eq:ode_reverse_process}
\end{equation}
Some recent works have found that ODE can achieve accelerated sampling compared to SDE, thereby significantly reducing the computational overhead of \textit{reverse denoising process}~\cite{ddim,dpm_solver,dpm_solver++}. In this paper, we implement accelerated sampling based on DPM Solver++. We derive the multiple condition-based classifier guidance based on~\cite{adm2021} in Sec.~\ref{sec:method}. Thus, our method is theoretically supported by this guidance, when compared to other GAN-based generation methods.
\section{Methodology}
\label{sec:method}
In this section, we first present the framework of DiffuseExpand, and then describe the processes involved in four stages (Stages I-IV).

\subsection{Total Framework.} 
As illustrated in Fig.~\ref{fig:overall_structure}, our DiffuseExpand is a four-stage algorithm for gradually improving the quality of the synthesized Image-Mask pairs and the alignment between the associated images and masks. Specifically, in Stage I, we fine-tune the pre-trained DPM without classifier guidance, so that it translates probability distributions from natural images to medical images. In Stage II, we train a segmenter for classifier guidance, this conditional guidance helps to finely tune the degree of matching of the image and mask in Stage III. We apply DPM Solver++ to complete the synthesis of Image-Mask pairs in Stage III. Lastly, in Stage IV, we discard the synthesized Image-Mask pairs that do not meet the criteria, which prescribes that $\mathcal{L}_{dice}$ exceeds a certain threshold. Based on this, the overall quality of the synthesized Image-Mask pairs is further improved.

\subsection{Stage I$:$ Fine-tune a pre-trained DPM}
Fine-tuning a DPM, pre-trained on ILSVRC2012, has shown to be a cost-effective and high-quality alternative for medical image segmentation~\cite{ILSVRC15}. In addition, the classifier-free approach is used to assist the training and enable more fine-grained conditional image synthesis. To accommodate both mask and image synthesis simultaneously, we perform fine-tuning on the model using multiple conditions in Stage I. Let us define a binary variable $y_1\sim p(y_1) =\mathcal{B}(0.5)$\footnote{Since there is the same number of masks and images in a dataset, the parameter for the Bernoulli distribution is set as $0.5$.} to indicate whether the DPM is synthesizing masks or images. More precisely, $y_1$ is $0/1$ means the DPM synthesizes the mask/image. We then define another variable $y_2$ that is dependent on $y_1$, where $p(y_2|y_1) = p(x^m)$, meaning $y_2$ represents masks and follows the distribution of the masks represented by $p(x^m)$. \textbf{For convenience, we follow the definitions of the symbols in Sec~\ref{sec:diffusion_model}}. As shown in Stage I of Fig.~\ref{fig:overall_structure}, we fine-tune the pre-trained model based on Mean Square Error (MSE), which can be specifically formulated as
\begin{equation}
\small
\begin{aligned}
& \mathcal{L}_\textrm{S\_I} = ||\widetilde{\mu}_{\theta} - \epsilon||_2^2 =\begin{cases}
	||\mu_\theta(x_t^i,t,y_1,x_0^m)-\epsilon||_2^2, & y_1 = 1,\\
	||\mu_\theta(x_t^m,t,y_1,\emptyset\ \ )-\epsilon||_2^2 , & y_1 = 0, \\
   \end{cases} \\ 
\end{aligned}
\label{eq:stage_1_loss}
\end{equation}
where $x^i_t$ and $x^m_t$ denote the image and mask at time $t$, respectively, with additional noise, \textit{i.e.}, ${x^i_t} = \alpha_t x_0^i + \sigma_t \epsilon, x^m_t=\alpha_t x_0^m + \sigma_t \epsilon$, where $x_0^i$ and $x_0^m$ represent the original image and mask, respectively, used in fine-tuning and $\emptyset$ refers to the corresponding input being empty. By fine-tuning the model using this loss function, the DPM is able to complete the two phases of sampling in Stage III and synthesize high-quality Image-Mask pairs. To be specific, we can synthesize $x_0^m$ by first making $y_1 = 0$, and then synthesize $x_0^i$ by making $y_1 = 1$ and conditioning on $y_2 = x_0^m$.

\subsection{Stage II$:$ Train Segmenter} To facilitate the sampling at Stage III with classifier guidance, we need to train a segmenter, which can be understood as a UNet-based classifier since it performs classification pixel by pixel at Stage II to model $p(y_1,y_2|x^i_t)$ and $p(y_1,y_2|x^m_t)$. We define the segmenter as $f_\psi$. Then, we leverage $\mathcal{L}_\textrm{S\_II}$ to accomplish two tasks: first, to discriminate whether inputs are images or masks, and second, to perform the segmentation task only when the input is an image:
\begin{equation}
\small
\begin{aligned}
 \mathcal{L}_\textrm{S\_{II}} =  \begin{cases}
& \mathcal{L}_{dice}(f_\psi(x^i_t),x^m_0)+\mathcal{L}_{bce}(f_\psi(x^i_t),x^m_0) +\\
&\quad\quad\mathcal{L}_{bce}(g_{\omega}(f_\psi(x^i_t)),y_1),\quad\quad\quad\quad  y_1 = 1,\\
&\mathcal{L}_{bce}(g_{\omega}(f_\psi(x^i_t)),y_1),\quad\quad\quad\quad\quad\quad  y_1 = 0,\\
   \end{cases} \\ 
\end{aligned}
\label{eq:stage_2_loss}
\end{equation}
where $\mathcal{L}_{dice}$ and $\mathcal{L}_{bce}$ denote Dice loss and Binary Cross Entropy loss, respectively, and $g_\omega$ refers to the additional encoder utilized for the prediction of $y_1$.

\subsection{Stage III$:$ Synthesize Image-Mask Pairs.} Here, we describe how to perform two phase-based sampling with the help of a fine-tuned DPM from Stage I and a trained segmenter from Stage II. Note that Stage I and Stage II are based on discrete \textit{forward diffusion process}.%, so that time is formulated as $t$. 
However, DPM Solver++ is designed based on ODE and therefore must convert discrete $t\in \{0,1,\cdots,T\}$ to continuous $\hat{t} \in [0,1]$ by $\hat{t} = t / T $. Next, we need to derive the sampling equation with multiple conditions so that the DPM can perform conditional image synthesis based on classifier guidance. We give the following Lemma~\ref{lem:ic_mi} and Lemma~\ref{lem:ic_nmi} (proof in Appendix~\ref{lem:ic_mi_appendix}).
\begin{lemma}
\label{lem:ic_mi}
For discrete reverse sampling $p_\theta(x_{t-1}|x_t) =\mathcal{N}(\mu_\theta(x_t,t),C)$, where $C$ is an arbitrary variance, if variables $y_{1:N}=\{y_1,\cdots y_N\}$ are mutually independent, then we can get $p_{\theta,\psi}(x_{t-1}|x_t,y_{1:N})\approx\mathcal{N}(x_{t-1};\hat{\mu}_{\theta,\psi}(x_t,t),C),$ where $ \hat{\mu}_{\theta,\psi}(x_t,t)=\mu_\theta(x_t,t)+C\left[\sum_{n=1:N}\nabla_{x_t}\log p_\psi(y_1|x_t)\right]$. And for continuous reverse sampling $dx_{\hat{t}} = \left[f({\hat{t}})x_{\hat{t}} - \frac{g^2({\hat{t}})}{2}\nabla_x \log p_{\theta}(x_{\hat{t}})\right]d{\hat{t}}$, if variables $y_{1:N}$ are mutually independent, then we can get $dx_{\hat{t}} = \left[f({\hat{t}})x_{\hat{t}} - \frac{g^2({\hat{t}})}{2}\nabla_x \log \hat{p}_{\theta,\psi}(x_{\hat{t}}|y_{1:N})\right]d{\hat{t}}$, where $\hat{p}_{\theta,\psi}(x_{\hat{t}}|y_{1:N}) =  {p}_{\theta}(x_{\hat{t}}) + \sum_{n=1:N}~{p}_{\psi}(y_n|x_{\hat{t}})$.
\end{lemma}
\begin{lemma}
\label{lem:ic_nmi}
For discrete inverse sampling $p_\theta(x_{t-1}|x_t) =\mathcal{N}(\mu_\theta(x_t,t),C)$, where $C$ is an arbitrary variance, if variables $y_{1:N}=\{y_1,\cdots y_N\}$ are mutually dependent, then we can get $p_{\theta,\psi}(x_{t-1}$ $|x_t,y_{1:N})\approx\mathcal{N}(x_{t-1};\hat{\mu}_{\theta,\psi}(x_t,t),C)$, where $\hat{\mu}_{\theta,\psi}(x_t,t)=\mu_\theta(x_t,t)+C[\nabla_{x_t}\log p_\psi(y_1|x_t)+\sum_{n=2:N}\nabla_{x_t}\log p_\psi(y_n|x_t,y_{1:n-1})]$. And for continuous reverse sampling $dx_{\hat{t}} = [f({\hat{t}})x_{\hat{t}} - \frac{g^2({\hat{t}})}{2}\nabla_x$ $\log p_{\theta}(x_{\hat{t}})]d{\hat{t}}$, if variables $y_{1:N}$ are mutually dependent, then we can get $dx_{\hat{t}} = [f({\hat{t}})x_{\hat{t}} - \frac{g^2({\hat{t}})}{2}\nabla_x$ $\log \hat{p}_{\theta,\psi}(x_{\hat{t}}|y_{1:N})]d{\hat{t}}$, where $\hat{p}_{\theta,\psi}(x_{\hat{t}}|y_{1:N}) =p_\psi(y_1|x_t)+\sum_{n=2:N}p_\psi(y_n|x_t,y_{1:n-1})$.
\end{lemma}
The lemmas we present are applicable to both independent/dependent conditional image synthesis with classifier guidance in discrete/continuous time. In our task, conditional image synthesis is performed based on continuous time $\hat{t}$ and with $y_1$ and $y_2$ are mutually dependent, thus achieving accelerated sampling by applying DPM Solver++. Therefore, our \textit{probability flow} ODE can be rewritten as ($\widetilde{\mu}_\theta$ is defined in Eq.~\ref{eq:stage_1_loss})
\begin{figure*}[t]
\centering
\begin{minipage}[t]{0.59\textwidth}
\includegraphics[width=1\linewidth,trim={0.5cm 3.2cm 1.2cm 2.3cm},clip]{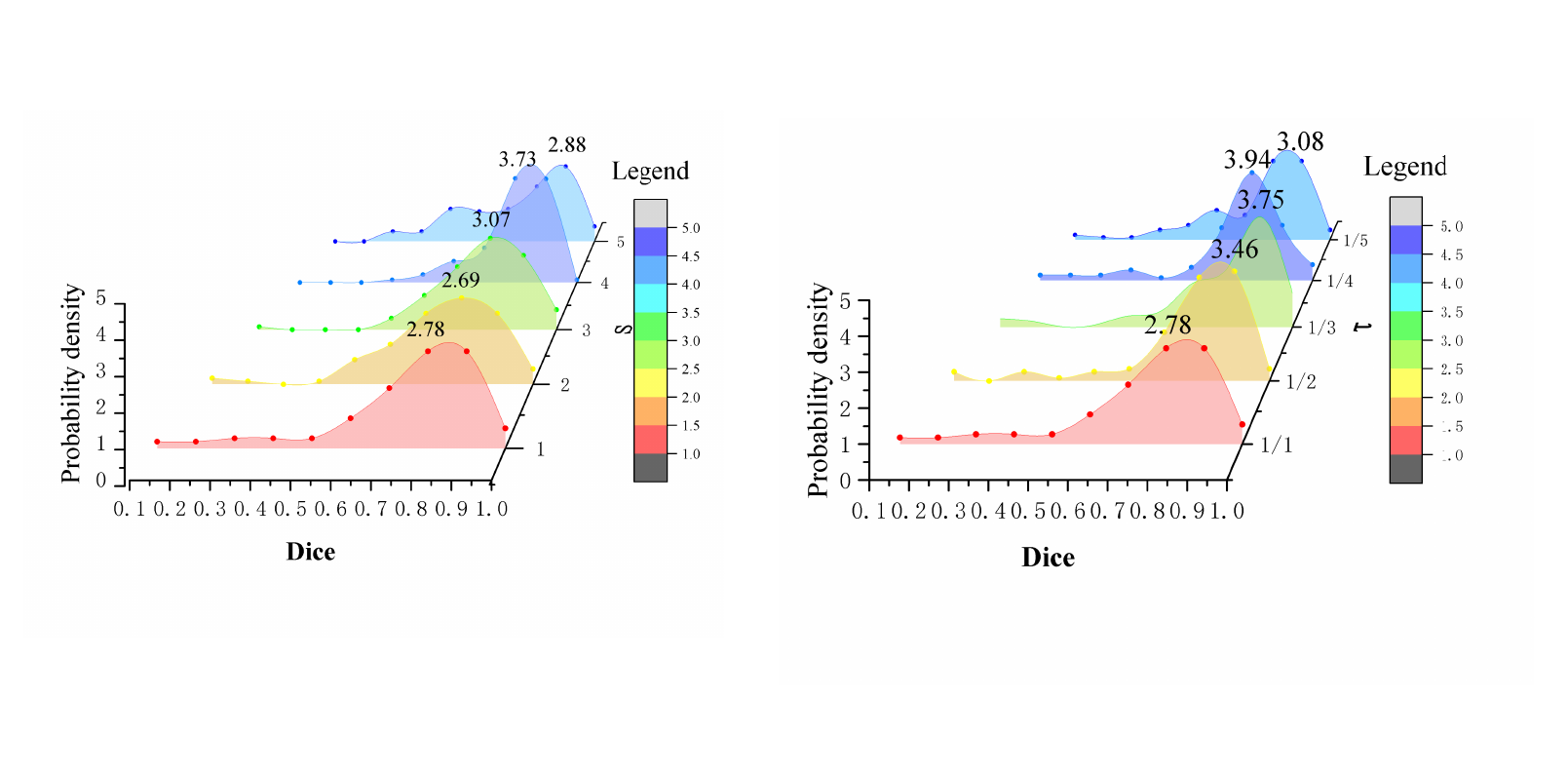}
\end{minipage}
\begin{minipage}[t]{0.39\textwidth}
\includegraphics[width=1\linewidth,trim={0cm 0cm 0cm 1.3cm},clip]{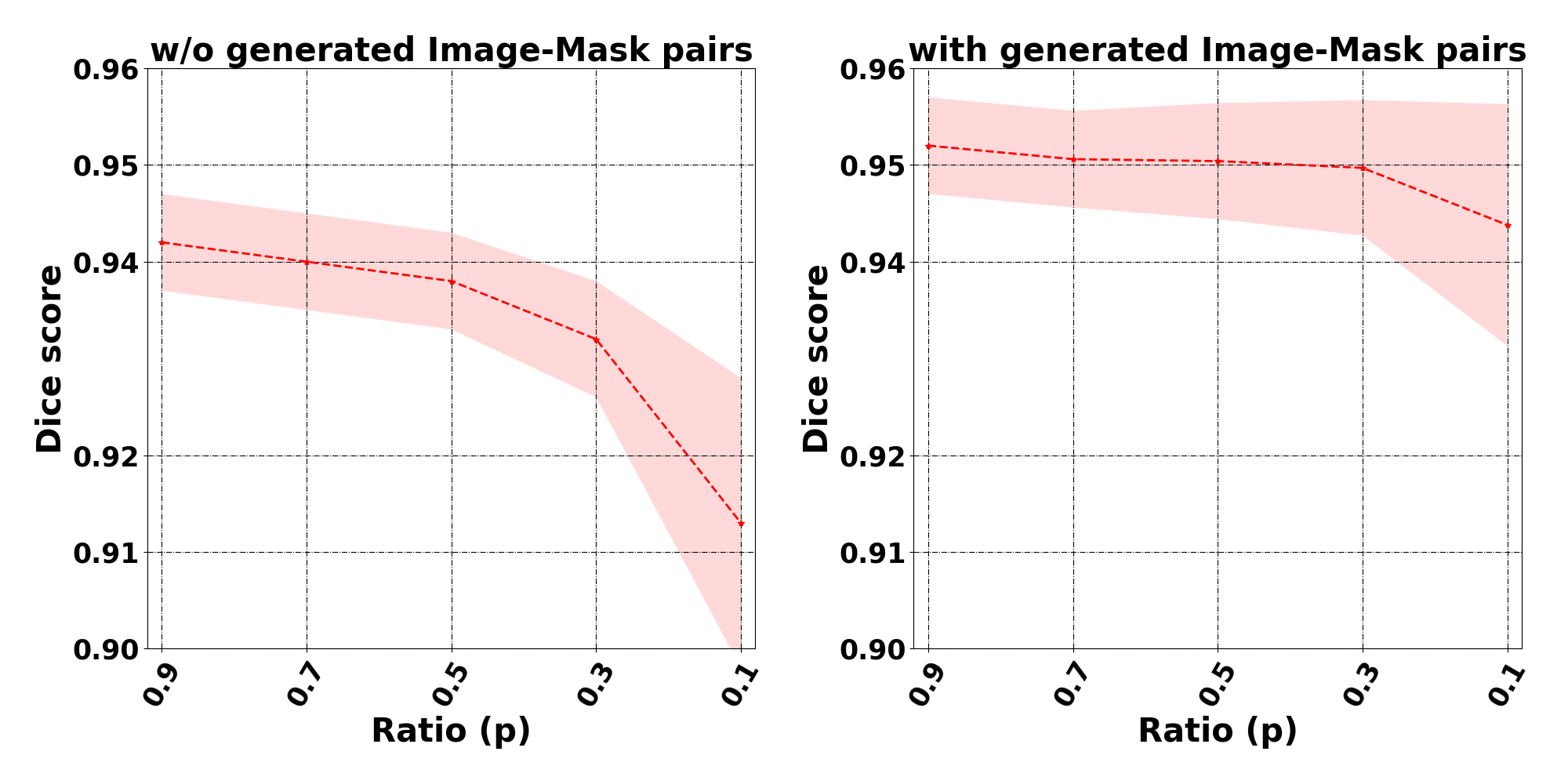}
\end{minipage}
\caption{\textbf{Left two:} Comparison experiments between $\tau$ and $s$ on CGMH Pelvis. We calculate the probability density of Dice Scores for samples obtained from DiffuseExpand, excluding Stage IV. These results are presented in a 3D waterfall plot. \textbf{Right two:} COVID-19 comparison experiments in few-shot scenarios. We employ 500 Image-Mask pairs synthesized by DiffuseExpand as the expanded dataset, following $\tau=1$. We use training sets with different split ratios, with the DPM and segmenter being retrained for each scenario to ensure fair comparisons.}
\label{fig:tau_and_few_shot}
\end{figure*}
\renewcommand\arraystretch{1.1}
\setlength\tabcolsep{3pt}%调列距
\begin{table*}[!t]
\center
\caption{$^+$: RandAugment is employed to translate training samples. $^*$: CutMix is made use of to translate training samples. origin: original training set. CF: Classifier Free. CG: Classifier Guidance. CS: Choose high-quality sample pairs based on Stage IV. }
\label{tab:ablation_study_result}
\resizebox{1.\textwidth}{!}{%
\begin{tabular}{l|l|rrrrrrrr}\toprule
\multirow{2}{*}{Dataset} & \multirow{2}{*}{Metric} & \multirow{2}{*}{Origin} & \multirow{2}{*}{CF} & \multirow{2}{*}{\begin{tabular}[c]{@{}c@{}}CF+CG \\ $\tau=1$\end{tabular}} &\multirow{2}{*}{\begin{tabular}[c]{@{}c@{}}CF+CG \\ $\tau=\frac{1}{2}$\end{tabular}}& \multirow{2}{*}{\begin{tabular}[c]{@{}c@{}}CF+CG \\ $\tau=\frac{1}{3}$\end{tabular}} & \multirow{2}{*}{\begin{tabular}[c]{@{}c@{}}CF+CG+CS \\ $\tau={1}$\end{tabular}} & \multirow{2}{*}{\begin{tabular}[c]{@{}c@{}}CF+CG+CS \\ $\tau=\frac{1}{2}$\end{tabular}} & \multirow{2}{*}{\begin{tabular}[c]{@{}c@{}}CF+CG+CS \\ $\tau=\frac{1}{3}$\end{tabular}} \\
& & & & & & & & & \\\hline
\multirow{5}{*}{\begin{tabular}[c]{@{}c@{}}COVID \\-19\end{tabular}} & FID$\downarrow$ & - & 9.100	 & 7.602 & 6.989 & 8.229 & 4.539 & \textbf{4.112} & 4.657\\
 & IS$\uparrow$ &1.009 & 1.008 & 1.008 & 1.008 & 1.008 & 1.009 & 1.009 & \textbf{1.010}\\\cline{2-10}
 & Dice$\uparrow$ &0.942$_{(\pm 0.005)}$ & 0.947$_{(\pm 0.005)}$& 0.943$_{(\pm 0.007)}$&0.946$_{(\pm 0.005)}$ & 0.946$_{(\pm 0.005)}$ & \textbf{0.952$_{(\pm 0.005)}$} &0.951$_{(\pm 0.005)}$ & \textbf{0.952$_{(\pm 0.005)}$} \\
& Dice$^+\!\uparrow$ &0.944$_{(\pm 0.005)}$&0.959$_{(\pm 0.002)}$&0.960$_{(\pm 0.002)}$&0.959$_{(\pm 0.003)}$&0.959$_{(\pm 0.001)}$&0.963$_{(\pm 0.001)}$&0.963$_{(\pm 0.002)}$& \textbf{0.965$_{(\pm 0.001)}$}\\
& Dice$^*\!\uparrow$ &0.931$_{(\pm 0.006)}$&0.943$_{(\pm 0.006)}$&0.946$_{(\pm 0.004)}$& \textbf{0.957$_{(\pm 0.007)}$} &0.945$_{(\pm 0.007)}$& 0.954$_{(\pm 0.004)}$ & 0.954$_{(\pm 0.003)}$ &0.952$_{(\pm 0.004)}$\\\hline
\multirow{5}{*}{\begin{tabular}[c]{@{}c@{}}CGMH \\Pelvis\end{tabular}} & FID$\downarrow$ & - & 2.825 &\textbf{2.535} & 2.657 & 2.736 & 6.413 & 7.205 & 7.592 \\
 & IS$\uparrow$ & \textbf{1.006} & 1.005 & \textbf{1.006} & 1.005 & 1.005 & \textbf{1.006} & \textbf{1.006} &\textbf{1.006} \\\cline{2-10}
 & Dice$\uparrow$ &0.941$_{(\pm 0.003)}$ &  0.931$_{(\pm 0.005)}$ & 0.946$_{(\pm 0.007)}$ & 0.948$_{(\pm 0.005)}$ & 0.951$_{(\pm 0.004)}$ & 0.962$_{(\pm 0.001)}$ & 0.962$_{(\pm 0.001)}$ & \textbf{0.963$_{(\pm 0.001)}$} \\
 & Dice$^+\!\uparrow$& 0.930$_{(\pm 0.008)}$&0.940$_{(\pm 0.006)}$&0.942$_{(\pm 0.008)}$&0.943$_{(\pm 0.007)}$&0.948$_{(\pm 0.004)}$& \textbf{0.966$_{(\pm 0.001)}$} & 0.964$_{(\pm 0.001)}$ & \textbf{0.966$_{(\pm 0.001)}$}\\
 & Dice$^*\!\uparrow$ &0.929$_{(\pm 0.006)}$ & 0.940$_{(\pm 0.006)}$&  0.943$_{(\pm 0.014)}$ & 
 0.942$_{(\pm 0.008}$ & 0.949$_{(\pm 0.006)}$ & 0.963$_{(\pm 0.001)}$ & 0.961$_{(\pm 0.004)}$ & \textbf{0.964$_{(\pm 0.002)}$}\\\bottomrule
\end{tabular}}
\end{table*}
\begin{figure*}[!t]
\centering
\includegraphics[width=0.95\linewidth,trim={0cm 0cm 0cm 0cm},clip]{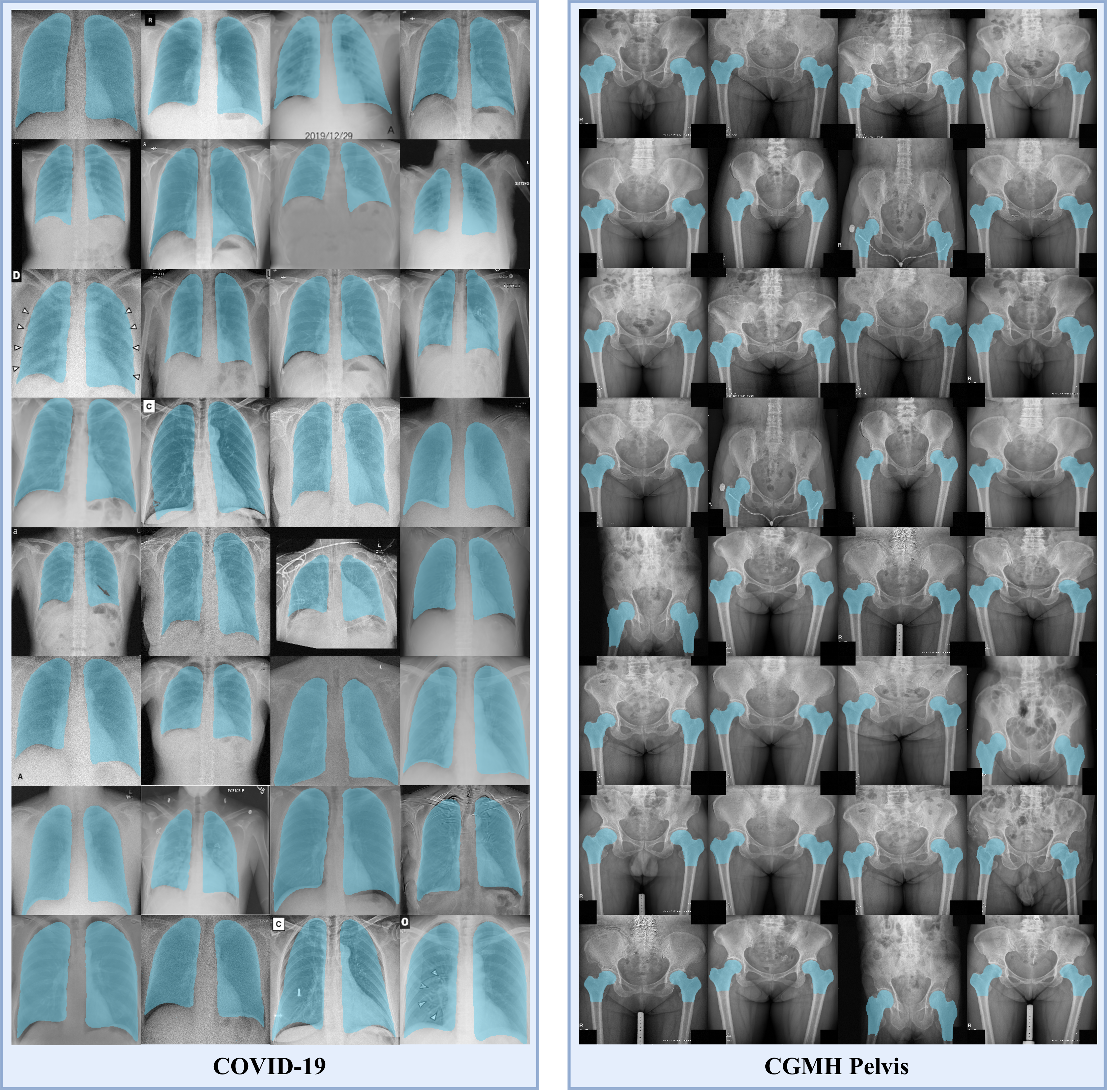}
\caption{Visualization of Image-Mask pairs synthesized by DiffuseExpand. We render the image and mask in one graph to better illustrate their correlation.}
\label{fig:visulaization}
\vspace{-8pt}
\end{figure*}
\begin{equation}
\begin{aligned}
& dx_{\hat{t}}\!=\![f(\hat{t})x_{\hat{t}}+g^2({\hat{t}})(\frac{\widetilde{\mu}_\theta}{2\sigma_{\hat{t}}} - \\
&\quad\quad\frac{1}{2}\nabla_x\log p_\psi(y_1|x_{\hat{t}}) + \frac{1}{2}\nabla_x\log p_\psi(y_2|x_{\hat{t}},y_1))]d{\hat{t}}.\\
\end{aligned}
\label{eq:stage_iii_process}
\end{equation}

Note that $x_{\hat{t}} \in \{x_{\hat{t}}^i,x_{\hat{t}}^m\}$. According to Eq.~\ref{eq:stage_iii_process}, we can first synthesize the mask by Gaussian noise and then synthesize the corresponding image by mask and Gaussian noise. In addition, Dhariwal \emph{et al.} propose to implement gradient scale $s$ to control the scale of $\nabla_x \log p_\psi(y|x_t)$, i.e., $s\nabla_x \log p_\psi(y|x_t)$. They interpret increasing $s$ as pushing $\left[\frac{p_\psi(\ve{y}_i|x_t)^s}{\sum_j p_\psi(\ve{y}_j|x_t)^s}\right]_{i}$ ($\ve{y}$ is the label vector) to a one-hot label. Unfortunately, Shu \emph{et al.}~\cite{shujian} point out that this interpretation is problematic because $\sum_j p_\psi(\ve{y}_j|x_t)^s$ is not a constant~\cite{adm2021}. To address this issue, we introduce the hyperparameter temperature $\tau$ often used in knowledge distillation~\cite{vanillakd}. Then, our novel \textit{score function} can then be denoted as $p(y|f_\psi(x_t)/\tau)$ and it is guaranteed that $\sum_j p(\ve{y}_j|f_\psi(x_t)/\tau)= 1$. Further, we give Lemma~\ref{lem:grad_softmax_tau} (proof in Appendix~\ref{lem:softmax_tau_2_appendix}) to reveal the association between $\tau$ and $s$.
\begin{lemma}
\label{lem:grad_softmax_tau}
For $p(y|f_\psi(x_{t}))$, if the normalized exponential function $p$ of the classifier is softmax or sigmoid, and $\frac{1}{\tau}=s\geq1$, then $\exists\ y_m$, satisfying $\min{(\{\ve{y}_j\}_{j})} \leq y_m \leq \max{(\{\ve{y}_j\}_{j})}$ ($\ve{y}$ is the logits of $f_\psi$) that makes the following inequality hold: $\left|\left|\frac{\partial y}{\partial x_t}\frac{\partial \log p(y|f_{\psi}(x_t))^s}{\partial y}\right|\right| \geq \left|\left|\frac{\partial y}{\partial x_t}\frac{\partial \log p(y|f_{\psi}(x_t)/\tau)}{\partial y}\right|\right|$ when $y \geq y_m$, $\left|\left|\frac{\partial y}{\partial x_t}\frac{\partial \log p(y|f_{\psi}(x_t))^s}{\partial y}\right|\right|$ $\leq \left|\left|\frac{\partial y}{\partial x_t}\frac{\partial \log p(y|f_{\psi}(x_t)/\tau)}{\partial y}\right|\right|$ when $y\leq y_m$. \end{lemma}
The variable $t$ here refers to either discrete or continuous time. Lemma~\ref{lem:grad_softmax_tau} reveals that $\tau$ has the ability to automatically adjust $||\frac{\partial y}{\partial x_t}||$ during conditional image synthesis. To be specific, the direction of the gradient $\frac{\partial y}{\partial x_t}$ is directed towards the target. If the larger the target $y$ is, which means that it is close to the target, then $\frac{\partial \log p(y|f_{\psi}(x_t)/\tau)}{\partial y}$ becomes smaller to avoid $\left|\left|\frac{\partial y}{\partial x_t}\frac{\partial \log p(y|f_{\psi}(x_t)/\tau)}{\partial y}\right|\right|$ being too large and skipping the target. If the target $y$ is smaller, $\tau$ has the ability to enlarge $||\frac{\partial y}{\partial x_t}||$.
\renewcommand\arraystretch{1.0}
\setlength\tabcolsep{3pt}%调列距
\begin{table*}[h]
\center
\caption{Comparative experiments for segmentation data expansion. All methods train the validation model only with their synthesized sample pairs.}
\label{tab:compare_results}
\resizebox{1.\textwidth}{!}{%
\begin{tabular}{l|rrr|r|rrr|r}\toprule
\multirow{3}{*}{Method} & \multicolumn{3}{c|}{COVID-19 (Dice$\uparrow$)} &  \multirow{3}{*}{\begin{tabular}[c]{@{}c@{}}COVID-19 \\ (FID$\downarrow$)\end{tabular}} & \multicolumn{3}{c|}{CGMH Pervis (Dice$\uparrow$)} &  \multirow{3}{*}{\begin{tabular}[c]{@{}c@{}}CGMH Pervis \\ (FID$\downarrow$)\end{tabular}}  \\
& \begin{tabular}[c]{@{}c@{}}UNet \\\cite{UNet} \end{tabular} & \begin{tabular}[c]{@{}c@{}}AttnUNet \\\cite{attnunet} \end{tabular}  & \begin{tabular}[c]{@{}c@{}}TransUNet \\\cite{transunet} \end{tabular} & & \begin{tabular}[c]{@{}c@{}}UNet \\\cite{UNet} \end{tabular} & \begin{tabular}[c]{@{}c@{}}AttnUNet \\\cite{attnunet} \end{tabular}  & \begin{tabular}[c]{@{}c@{}}TransUNet \\\cite{transunet} \end{tabular}& \\\midrule
Synthmed & 0.569$_{(\pm 0.157)}$ & 0.510$_{(\pm 0.147)}$ &  0.744$_{(\pm 0.024)}$ & 32.353 & 0.532$_{(\pm 0.178)}$& 0.581$_{(\pm 0.162)}$& 0.767$_{(\pm 0.131)}$& 38.472 \\
XLsor & 0.944$_{(\pm 0.003)}$& 0.947$_{(\pm 0.005)}$ & 0.925$_{(\pm 0.006)}$ & \textbf{1.197} & 0.924$_{(\pm 0.003)}$& 0.935$_{(\pm 0.004)}$ & 0.881$_{(\pm 0.103)}$ &  \textbf{1.079} \\
DiffuseExpand & \textbf{0.948$_{(\pm 0.002)}$} & \textbf{0.949$_{(\pm 0.003)}$} & \textbf{0.948$_{(\pm 0.003)}$} &  4.657 & \textbf{0.930$_{(\pm 0.005)}$} & \textbf{0.936$_{(\pm 0.014)}$} &\textbf{ 0.950$_{(\pm 0.002)}$} & 7.592 \\\bottomrule
\end{tabular}}
\end{table*}

\subsection{Stage IV$:$ Choose high-quality Image-Mask pairs} Generate models always synthesize some bad samples, and Stage IV exists to discard the bad Image-Mask pairs synthesized in Stage III. We utilize the neural network to eliminate the bad Image-Mask pairs and retain the high-quality ones. Specifically, by fixing a well-trained neural network and then calculating $\mathcal{L}_{dice}$ from a given synthesized Image-Mask pair, the sample is kept if it is less than a certain threshold $\eta$; otherwise, the pair is discarded. This approach removes the image and mask misaligns in a pair or poorly synthesized pairs.
\section{Experiment}
Below, we first introduce the relevant datasets and hyperparameter settings used in our experiments, followed by the presentation of qualitative and quantitative experimental ablation studies. Then we present the results of our comparative trial. Finally, we show the results of the visualization of the synthesized sample pairs.
\subsection{Settings} We verify the efficacy of DiffuseExpand on two distinct 2D X-ray datasets, namely COVID-19~\cite{covid} and CGMH Pelvis~\cite{cgmh}. The COVID-19 dataset comprises chest X-rays, while the CGMH Pelvis dataset contains pelvic X-rays. Notably, both datasets encompass 304 and 400 Image-Mask pairs, correspondingly. By default, we have divided the training and testing sets into a ratio of 9 to 1. Our DiffuseExpand is trained solely on the training set. Moreover, to fine-tune the DPM and train the segmenter, we set the batch size to 16, the number of iterations to 30,000, the optimizer to Adam, and the learning rate to 1e-4. Subsequently, we expand the training data by incorporating the synthesized Image-Mask pairs to train the validated model and compute its Dice Score. During this stage, we used a batch size of 16, trained for 50 epochs, and employed Adam with a learning rate of 1e-2. In particular, for Stage IV we use 6.5e-2 as $\eta$ in the expectation that the quality of the sample pairs can be effectively improved by rigorous screening. All experiments that utilize the validated model to compute Dice Score are repeated three times.

\subsection{Ablation Study} We partition the core algorithm of DiffuseExpand into three parts, namely Classifier Free, Classifier Guidance, and Choose high-quality sample pairs in Stage IV. Classifier Free refers to synthesizing Image-Mask pairs in Stage III solely using the DPM to embed conditions, without the assistance of the segmenter. On the other hand, Classifier Guidance involves conditional synthesis in Stage III with the aid of the segmenter. To investigate the impact of these parts on the quality (measured by Inception Score, IS~\cite{is}) and diversity (measured by Frechet Inception Distance, FID~\cite{fid}) of the synthesized images~\cite{adm2021}, and to evaluate whether the synthesized Image-Mask pairs, as an expanded dataset, can enhance the model's (UNet~\cite{UNet}) performance (measured by Dice Score), we conduct ablation experiments and present the results in Table~\ref{tab:ablation_study_result}. These experiments, conducted in various data augmentation scenarios, confirm the effectiveness of the aforementioned three parts. The results also demonstrate that a slight reduction in $\tau$ reduces FID and enhances the generalization ability of the validated model. Since Stage IV only retains the sample pairs with highly aligned masks and images, which seriously inhibits the synthesized sample pairs' diversity, changes in $\tau$ have little effect on the Dice Score of the validated model. We compare the validity of $\tau$ in the Classifier Guidance and $s$ and present the results in Fig.~\ref{fig:tau_and_few_shot} (left). For each scenario, we use 100 Image-Mask pairs to obtain the statistics. The results indicate that Image-Mask pairs synthesized by gradient scaling with $\tau$ yield higher Dice Score values than those synthesized with $s$ when $\frac{1}{\tau} = s$. Therefore, using $\tau$ to guide image synthesis can be considered more effective than using $s$ in certain cases. Finally, we aim to verify the effectiveness of DiffuseExpand in \textbf{few-shot scenarios}. With the training approach outlined in the notes of Fig.\ref{fig:tau_and_few_shot} (right), we observe that the model's performance is improved significantly when training with both the original and synthesized sample pairs compared to training with only the original sample pairs on a smaller dataset. 

\subsection{Comparison Results} Given the limited research on segmentation data expansion and the lack of open-source code, we compare two algorithms for this task: Synthmed~\cite{compare_1} and XLsor~\cite{compre_2}. These algorithms use conditional GANs and manual data augmentation, respectively, to expand the data. The comparison results are presented in Table~\ref{tab:compare_results}, which shows that DiffuseExpand outperforms Synthmed and XLsor in terms of Dice Score. Notably, XLsor is not a generative algorithm in some sense because it only augments the original samples. The difference in distribution between the samples obtained by XLsor and the original samples is very small. Thus, XLsor achieves a better FID than DiffuseExpand.

\subsection{Visualization} The synthesized results of DiffuseExpand are presented in Fig.~\ref{fig:visulaization}. It is evident that all synthesized samples exhibit both high quality and diversity, with their corresponding image masks accurately labeled. Notably, the DiffuseExpand method is capable of restoring special details, such as the sample displayed in top 1, left 3 (COVID-19), featuring the date "2019/12/29", which serves as further confirmation of the outstanding performance of DiffuseExpand.
\section{Conclusion}
In this paper, we propose DiffuseExpand, a four-stage conditional synthesis algorithm based on DPMs to expand the segmentation dataset. Experiments show that DiffuseExpand can synthesize high-quality Image-Mask pairs, which demonstrates the feasibility of applying DPMs for segmentation data expansion. In future research, we aim to improve the conditional image synthesis process to generate highly aligned Image-Mask pairs without requiring Stage IV.
\appendix
\section{Appendix}
\begin{proof}
\label{lem:ic_mi_appendix}
Lemma~\ref{lem:ic_mi}. For discrete reverse sampling, with Bayes' theorem, we get
\begin{equation}
\begin{aligned}
& p_\psi(x_{t-1}|y_{1:N}) =  \frac{p_\psi(y_{1:N}|x_{t-1})p(x_{t-1})}{p(y_{1:N})}. \\
\end{aligned}
\label{eq:bayes}
\end{equation}
Conditional on the more noisy sample $x_t$ and using the difference approximation, the sampling equation can be written as
\begin{equation}
\begin{aligned}
&\quad\log p_{\theta,\psi}(x_{t-1}|x_t,y_{1:N}) \\
&= \log p(x_{t-1}|x_t)\!+\!\log p_\psi(y_1|x_{t-1})\!+\!\cdots\!+\!\log p_\psi(y_N|x_{t-1}) \\
&\quad - \!\log p_\psi(y_1|x_{t})\!-\!\cdots\!-\!\log p_\psi(y_N|x_{t}) \\
& \approx \log p(x_{t-1}|x_t) + (x_{t-1}-x_t)[\nabla_{x_t}\log p_\psi(y_1|x_{t})\\
&\quad + \cdots\!+\!\nabla_{x_t}\log p_\psi(y_N|x_{t})] \\
& = -||x_{t-1}-\mu_\theta(x_t,t)||^2/2C+ (x_{t-1}-x_t)\\
&\quad [\nabla_{x_t}\log p_\psi(y_1|x_{t})+\cdots+\!\nabla_{x_t}\log p_\psi(y_N|x_{t})] + f_1(x_t) \\
& =\frac{-1}{2C }[x^T_{t-1}x_{t-1} + x_{t-1}(-2\mu_\theta(x_t,t)^T \\
&\quad +2C\nabla_{x_t}(\log p_\psi(y_1|x_{t})\!-\!\cdots\!-\!\log p_\psi(y_N|x_{t})))+f_2(x_t)] \\
\end{aligned}
\nonumber
\end{equation}
\begin{equation}
\begin{aligned}
& = \frac{-1}{2C}||x_{t-1} - [\underbrace{\mu_\theta(x_t,t)+C[\sum_{n=1:N}\nabla_{x_t}\log p_\psi(y_n|x_{t})]}_{\hat{\mu}_{\theta,\psi}(x_t,y_{1:N},t)}]||^2\\
&\quad+f_3(x_t),\\
\end{aligned}
\label{eq:proof_1}
\end{equation}
where $f_1,f_2,f_3$ are functions that are independent of $x_{t-1}$. Thus, we can derive the posterior probability distribution $p_{\theta,\psi}(x_{t-1}|x_t,y_{1:N})$ as $\mathcal{N}(x_{t-1};\hat{\mu}_{\theta,\psi}(x_t,y_{1:N},t),C)$.

In addition, for continuous reverse sampling, we know that $\nabla_x \log p_\psi(x|y_{1:N}) = \nabla_x \log p(x) + \nabla_x \log p_\psi(y_{1:N}|x)$. Because $y_{1:N}$ are mutually independent, it is possible to obtain $dx_{\hat{t}} = \left[f(t)x_{\hat{t}} - \frac{g^2({\hat{t}})}{2}\nabla_x \log \hat{p}_{\theta,\psi}(x_{\hat{t}}|y_{1:N})\right]d{\hat{t}},$ \textrm{where} $\hat{p}_{\theta,\psi}(x_{\hat{t}}|y_{1:N}) =  {p}_{\theta}(x_{\hat{t}}) + \sum_{n=1:N}{p}_{\psi}(y_n|x_{\hat{t}})$.
\end{proof}
\begin{proof}
\label{lem:ic_nmi_appendix}
Lemma~\ref{lem:ic_nmi}. From Lemma~\ref{lem:ic_mi} and the formula $p(y_1)p(y_2|y_1)\cdots p(y_N|y_{1:N-1}) = p(y_{1:N})$, we can conclude that Lemma~\ref{lem:ic_nmi} holds.
\end{proof}

\begin{proof}
\label{lem:softmax_tau_2_appendix}
Lemma~\ref{lem:grad_softmax_tau}.
\textbf{The normalized exponential function $p$ of the classifier is softmax.} We denote $f_\psi(x_{t})$ as a vector $\ve{y}$ of length $C$, with the target $y$ denoted as $\ve{y}_i$, then we calculate the partial derivatives with respect to $x_t$ for $\log p(y|f_\psi(x_{t}))^s$ and $\log p(y|f_\psi(x_{t})/\tau)$ respectively.
\begin{equation}
\begin{aligned}
& \frac{\partial \log p(y|f_{\psi}(x_t))^s}{\partial x_t} = \frac{\partial \log p(y|f_{\psi}(x_t))^s}{\partial \ve{y}_i}\frac{\partial \ve{y}_i}{\partial x_t}\\
&\quad +\sum_{k\neq i} \frac{\partial \log p(y|f_{\psi}(x_t))^s}{\partial \ve{y}_k}\frac{\partial \ve{y}_k}{\partial x_t} \\
& =  \frac{\partial \ve{y}_i}{\partial x_t} \frac{\partial \log \left(\frac{e^{\ve{y}_i}}{\sum_j e^{\ve{y}_j}}\right)^s}{\partial \ve{y}_i} + \sum_{k\neq i}  \frac{\partial \ve{y}_k}{\partial x_t}\left[-s\frac{e^{\ve{y}_k}}{\sum_j e^{\ve{y}_j}}\right] \\
& =  \frac{\partial \ve{y}_i}{\partial x_t} \frac{s\left(\frac{e^{\ve{y}_i}}{\sum_j e^{\ve{y}_j}}\right)^{s-1}\left[\frac{e^{\ve{y}_i}}{\sum_j e^{\ve{y}_j}} - \left(\frac{e^{\ve{y}_i}}{\sum_j e^{\ve{y}_j}}\right)^2\right]}{\left(\frac{e^{\ve{y}_i}}{\sum_j e^{\ve{y}_j}}\right)^s} \\
&\quad + \sum_{k\neq i}  \frac{\partial \ve{y}_k}{\partial x_t}\left[-s\frac{e^{\ve{y}_k}}{\sum_j e^{\ve{y}_j}}\right] \\
& =  \frac{\partial \ve{y}_i}{\partial x_t} \left[s\left(1-\left(\frac{e^{\ve{y}_i}}{\sum_j e^{\ve{y}_j}}\right)\right)\right]  + \sum_{k\neq i}  \frac{\partial \ve{y}_k}{\partial x_t}\left[-s\frac{e^{\ve{y}_k}}{\sum_j e^{\ve{y}_j}}\right] .\\
\end{aligned}
\end{equation}
\begin{equation}
\begin{aligned}
&  \frac{\partial \log p(y|f_{\psi}(x_t)/\tau)}{\partial x_t}  = \frac{\partial \log p(y|f_{\psi}(x_t)/\tau)}{\partial \ve{y}_i}  \frac{\partial \ve{y}_i}{\partial x_t}  \\
&\quad + \sum_{k\neq i} \frac{\partial \log p(y|f_{\psi}(x_t)/\tau)}{\partial \ve{y}_k}\frac{\partial \ve{y}_k}{\partial x_t} \\
& =  \frac{\partial \ve{y}_i}{\partial x_t} \frac{\partial  \log\left(\frac{e^{\ve{y}_i/\tau}}{\sum_j e^{\ve{y}_j/\tau}}\right)}{\partial \ve{y}_i}+ \sum_{k\neq i}  \frac{\partial \ve{y}_k}{\partial x_t}\left[-\frac{1}{\tau}\frac{e^{\ve{y}_k/\tau}}{\sum_j e^{\ve{y}_j/\tau}}\right] \\
\end{aligned}
\nonumber
\end{equation}
\begin{equation}
\begin{aligned}
& =  \frac{\partial \ve{y}_i}{\partial x_t} \frac{\frac{1}{\tau}\left[\frac{e^{\ve{y}_i/\tau}}{\sum_j e^{\ve{y}_j/\tau}} - \left(\frac{e^{\ve{y}_i/\tau}}{\sum_j e^{\ve{y}_j/\tau}}\right)^2\right]}{\left(\frac{e^{\ve{y}_i}/\tau}{\sum_j e^{\ve{y}_j/\tau}}\right)} \\
&\quad + \sum_{k\neq i}  \frac{\partial \ve{y}_k}{\partial x_t}\left[-\frac{1}{\tau}\frac{e^{\ve{y}_k/\tau}}{\sum_j e^{\ve{y}_j/\tau}}\right]  \\
&  =  \frac{\partial \ve{y}_i}{\partial x_t} \left[\frac{1}{\tau}\left(1-\left(\frac{e^{\ve{y}_i/\tau}}{\sum_j e^{\ve{y}_j/\tau}}\right)\right)\right] + \sum_{k\neq i}  \frac{\partial \ve{y}_k}{\partial x_t}\left[-\frac{1}{\tau}\frac{e^{\ve{y}_k/\tau}}{\sum_j e^{\ve{y}_j/\tau}}\right] . \\
\end{aligned}
\label{eq:grad_softmax_tau}
\end{equation}
Then we want to know how $\frac{e^{\ve{y}_i/\tau}}{\sum_j e^{\ve{y}_j/\tau}}$ varies due to changes in $\frac{1}{\tau}$. By solving for the partial derivative with respect to $\frac{1}{\tau}$ we obtain
\begin{equation}
\small
\begin{aligned}
&\frac{\partial \frac{e^{\ve{y}_i/\tau}}{\sum_j e^{\ve{y}_j/\tau}}}{\partial \frac{1}{\tau}} =  \frac{e^{2\ve{y}_i+\ve{y}_i/\tau}}{(\sum_j e^{\ve{y}_j/\tau})^2}\left[\sum_j (\ve{y}_i - \ve{y}_j) e^{(\ve{y}_i-\ve{y}_j)}\right]\\
\end{aligned}
\label{eq:grad_softmax_tau_2}
\end{equation}
When $\frac{1}{\tau} \geq 1$ and $\sum_j (\ve{y}_i - \ve{y}_j) e^{(\ve{y}_i-\ve{y}_j)}\geq 0$, then $\frac{e^{\ve{y}_i/\tau}}{\sum_j e^{\ve{y}_j/\tau}}\geq \frac{e^{\ve{y}_i}}{\sum_j e^{\ve{y}_j}}$. And when $\frac{1}{\tau} \geq 1$ and $\sum_j (\ve{y}_i - \ve{y}_j) e^{(\ve{y}_i-\ve{y}_j)} \leq 0 $, then $\frac{e^{\ve{y}_i/\tau}}{\sum_j e^{\ve{y}_j/\tau}}\leq \frac{e^{\ve{y}_i}}{\sum_j e^{\ve{y}_j}}$. After that, we can discover that the function $g(x) = \sum_j (x - \ve{y}_j) e^{(x-\ve{y}_j)}$ is a monotonically increasing function. There exists a value $y_m$ that $\min{(\{\ve{y}_j\}_{j=1}^C)} \leq y_m \leq \max{(\{\ve{y}_j\}_{j=1}^C)}$, satisfying $g(y_m)=0$. Therefore, we can prove that
\begin{equation}
\small
\begin{aligned}
& \left|\left|\frac{\partial \ve{y}_i}{\partial x_t}\frac{\partial \log p(y|f_{\psi}(x_t))^s}{\partial \ve{y}_i}\right|\right| \geq \left|\left|\frac{\partial \ve{y}_i}{\partial x_t}\frac{\partial \log p(y|f_{\psi}(x_t)/\tau)}{\partial \ve{y}_i}\right|\right|\\
&\quad\quad\quad\quad\quad\quad\quad\quad\quad\quad\quad\quad\quad\quad\quad\quad,\frac{1}{\tau}=s \geq 1\ \textrm{and}\ \ve{y}_i\geq y_m,\\
& \left|\left|\frac{\partial \ve{y}_i}{\partial x_t}\frac{\partial \log p(y|f_{\psi}(x_t))^s}{\partial \ve{y}_i}\right|\right| \leq \left|\left|\frac{\partial \ve{y}_i}{\partial x_t}\frac{\partial \log p(y|f_{\psi}(x_t)/\tau)}{\partial \ve{y}_i}\right|\right|\\
&\quad\quad\quad\quad\quad\quad\quad\quad\quad\quad\quad\quad\quad\quad\quad\quad,\frac{1}{\tau}=s \geq 1\ \textrm{and}\ \ve{y}_i\leq y_m,\\
\end{aligned}
\label{eq:grad_softmax_tau_2}
\end{equation}
\textbf{The normalized exponential function $p$ of the classifier is sigmoid.} In this case, $p(y|f_\psi(x_{t}))$ is a value. Also, $p(y|f_\psi(x_{t}))$ and $1 -p(y|f_\psi(x_{t}))$ represent the probability of classification into two classes corresponding to each other. Denote $f_\psi(x_{t})$ as $\ve{y}_0$, and then think about the vector $[0,\ve{y}_0]^T$. We let $\ve{y}_0$ be the input to the sigmoid. Thus, $p(y|f_\psi(x_{t}))$ can be formulated as
\begin{equation}
\small
\begin{aligned}
&\textrm{sigmoid}(\ve{y}_0) = \frac{1}{1+e^{-\ve{y}_0}} = \frac{e^{\ve{y}_0}}{e^0+e^{\ve{y}_0}} = \textrm{softmax}([0,\ve{y}_0]^T).\\
\end{aligned}
\label{eq:sigmoid_tau_1}
\end{equation}
Therefore, the case when $p$ is sigmoid can be interpreted in terms of the case when $p$ is softmax.
\end{proof}

\clearpage

{\small
\bibliographystyle{named}
\bibliography{egbib}
}

\end{document}